\documentclass[twocolumn,aps,amsmath,amssymb,prl,footinbib]{revtex4}
\usepackage{graphicx}
\usepackage{bm}         
\usepackage{color}         
\usepackage{amsmath}
\usepackage{units}
\usepackage{bbm}

\date{\today}
\begin{document}

\title{Majorana-Weyl crossings in topological multi-terminal junctions}
\author{Manuel~Houzet and Julia~S.~Meyer}
\affiliation{Univ.~Grenoble Alpes, CEA, INAC-{Pheliqs}, F-38000 Grenoble, France}

\begin{abstract}
We analyze the Andreev spectrum in a four-terminal Josephson junction between one-dimensional topological superconductors in class D. We find that a topologically protected crossing in the space of three superconducting phase differences can occur between the two lowest Andreev bound states. This crossing can be detected through the transconductance quantization, in units of $2e^2/h$, between two voltage-biased terminals. Our prediction provides yet another example of topology in multi-terminal Josephson junctions. We discuss possible realizations of such junctions with semiconducting crossed nanowires and with quantum-spin Hall insulators.
\end{abstract}

\maketitle

{\em Introduction.} 
It was long realized that an arbitrary Hamiltonian parametrically controlled by three parameters admits for topologically protected crossings in its energy spectrum~\cite{Herring1937}. In the vicinity of a crossing, the Hamiltonian in this three-parameter space takes the same form as the one introduced by Hermann Weyl in 1929 to describe relativistic massless particles in three dimensions. The crossings are now called Weyl points. This finding, as well as its generalization to physical systems protected by additional symmetries, was important for the prediction of topological properties in various areas of condensed matter, optical, or mechanical physics. 

In a recent work~\cite{Riwar2016}, it was predicted that such Weyl crossings appear at zero energy in the Andreev spectrum of four-terminal Josephson junctions made with conventional $s$-wave superconductors connected through a common normal scattering region. In that case, the three parameters are three superconducting phase differences between the four leads. Due to spin-rotation symmetry, such crossings are the only allowed states at zero energy. At such \lq\lq Andreev-Weyl\rq\rq\ crossings, the Chern number of the ground state in a submanifold of the phase space characterized by two phase differences changes. As a consequence,
the Andreev-Weyl crossings would manifest themselves through a quantized transconductance, in units of $4e^2/h$, between two voltage-biased terminals~\cite{Riwar2016,Eriksson2017}. This prediction was subsequently extended to junctions with three terminals in an external magnetic field~\cite{Meyer2017,Levchenko2017}.

On the other hand, Andreev-Weyl crossings have been shown to shift away from zero-energy, if spin-rotation symmetry is broken due to, {\it e.g.},  spin-orbit coupling~\cite{Yokoyama2015}. As long as the shift is small, the lowest energy state will cross the Fermi level on a surface surrounding the Weyl point in the space of the three phases. The prediction for the transconductance quantization away from the Weyl points remains valid. On the other hand, spin-orbit coupling may lead to the appearance of topological superconductivity with Majorana edge states~\cite{Read2000,Kitaev2001,Fu2008}. Possible realizations using semiconductor nanowires~\cite{Lutchyn2010,Oreg2010} are studied extensively. Four-terminal junctions based on the same kind of materials have already been realized, and one may wonder whether they might have similar properties to those described above. In this paper, we show that indeed they do, but with a significant twist compared to the previously studied case. The lowest Andreev state in such junctions depends $4\pi$-periodically~\cite{Kitaev2001} on the superconducting phase differences, which is a hallmark of the Majorana physics. As such it crosses the Fermi level along surfaces in the three-parameter phase space of a four-terminal junction. However, it can have finite-energy Weyl crossings with the next Andreev level, which is -- by contrast -- $2\pi$-periodic. Below we show that these \lq\lq Majorana-Weyl\rq\rq~crossings occur with a large probability in specific models. In the presence of such crossings, the $2\pi$-periodic state acquires a finite Chern number. As before, a finite Chern number is associated with a quantized transconductance, but now in units of $2e^2/h$ due to the lifted spin degeneracy. Our predictions could be tested with the platforms of semiconducting nanowires~\cite{Gazibegovic2017} and heterostructures~\cite{Suominen2017} that are currently investigated for the detection and manipulation of Majorana modes.

\begin{figure}
\includegraphics[width=0.7\columnwidth]{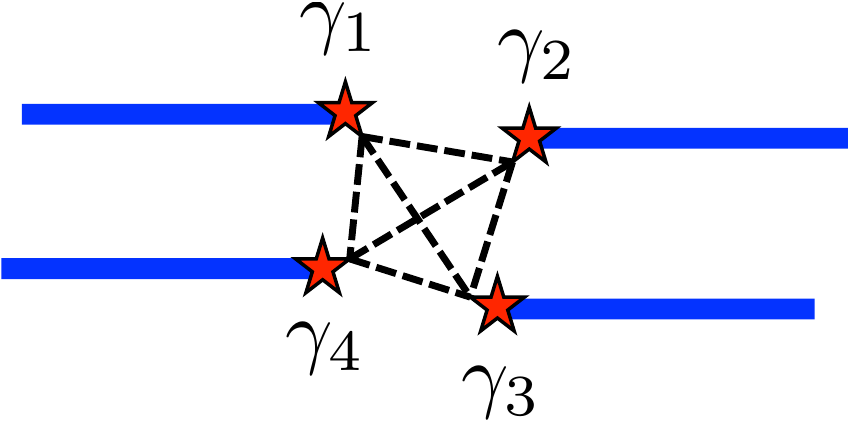}
\caption{
\label{fig:setup} 
Four-terminal junction formed of one-dimensional topological superconducting leads (blue lines) accommodating Majorana zero modes $\gamma_i$ at their extremities (red stars), and in the presence of a weak tunnel coupling between any pair of leads (dashed lines).
}
\end{figure} 

{\em Tunnel junction.} The physics can be most easily understood in the case of a tunnel junction made of four weakly coupled one-dimensional spinless $p$-wave superconductors~\cite{Kitaev2001}, corresponding to class D in the classification of topological insulators and superconductors~\cite{Chiu2016}, as illustrated in Fig.~\ref{fig:setup}. In that case, the effective low-energy Hamiltonian can be written in terms of the four Majorana end modes at the junction. It takes the form
\begin{equation}
\label{eq:Htunnel}
H=\frac i2\sum_{1\leq a<b\leq 4} \xi_{ab} \gamma_a\gamma_b
\end{equation}
with
\begin{equation}
\label{eq:xi}
\xi_{ab}= |t_{ab}|\sin\left(\frac{\chi_a-\chi_b}2-\phi_{ab}\right).
\end{equation}
Here, $\gamma_a$ is a Majorana fermionic operator (such that $\gamma_a^2={1}$), which describes the Majorana zero mode at the end of superconductor $a$, with superconducting phase  $\chi_a$~\cite{footnote-phase}, when it is decoupled from the others. (Without loss of generality we set $\chi_4=0$ below.) Furthermore, $t_{ab}=|t_{ab}| e^{i\phi_{ab}}$ are proportional to the tunneling matrix elements for electrons between leads $a$ and $b$. Note that Hermiticity of the %normal 
tunnel Hamiltonian requires $t_{ba}^{}= t_{ab}^*$. %(In particular, $t_{aa}$ is real.)  
The Hamiltonian \eqref{eq:Htunnel} accounts for all (bound) states in an energy bandwidth $\ll \Delta$, where $\Delta$ is the superconducting gap, provided that the transmission probabilities between the leads are small.

Next, we introduce fermion operators, $c_+={(\gamma_1+i\gamma_2)/2}$ and $c_-={(\gamma_3+i\gamma_4)/2}$. Then, Hamiltonian~\eqref{eq:Htunnel} reads $H=\frac 12 {\cal C}^\dagger {\cal H}{\cal C}$, where ${\cal C}=(c_+,c_-,c_-^\dagger,-c_+^\dagger)^T$ is an annihilation/creation operator in particle/hole space and
\begin{equation}
\label{eq:BdG}
{\cal H}=\bm{S}\cdot\bm{\sigma}+\bm{T}\cdot\bm{\tau}
\end{equation}
is a Bogoliubov-de~Gennes (BdG) Hamiltonian. Here $\bm{\sigma}=(\sigma_x,\sigma_y,\sigma_z)$ and $\bm{\tau}=(\tau_x,\tau_y,\tau_z)$ are vectors of Pauli matrices in $(c_+,c_-)$-space and $(c,c^\dagger)$-space, respectively, and
\begin{subequations}
\label{eq:ST}
\begin{eqnarray}
\bm{S}&=&(\xi_{14}-\xi_{23},-\xi_{13}-\xi_{24},\xi_{12}-\xi_{34}),\\
\bm{T}&=&(-\xi_{14}-\xi_{23},-\xi_{13}+\xi_{24},\xi_{12}+\xi_{34}).
\end{eqnarray}
\end{subequations}
Hamiltonian \eqref{eq:BdG}, which is the sum of two Weyl Hamiltonians, admits for four eigenstates with pairwise opposite energies, $E_{\sigma\tau}=\sigma\left(|\bm{S}|+\tau|\bm{T}|\right)$ with $\sigma,\tau=\pm$. 
We readily check with Eqs.~\eqref{eq:xi} and \eqref{eq:ST} that $E_{\sigma+}$ and $E_{\sigma-}$ depend $2\pi$- and $4\pi$-periodically on the phase differences, respectively. Namely, $E_{\sigma\pm}(\chi_1+2\pi,\chi_2,\chi_3)=\pm E_{\sigma\pm}(\chi_1,\chi_2,\chi_3)$ and similar relations with the other phases~\cite{footnote}. This is expected as $|\sigma+\rangle$ is a conventional Andreev bound state whose energy remains above (if $\sigma=+$) or below (if $\sigma=-$) the Fermi level at all phases, while $|\sigma-\rangle$ is a Majorana-Andreev bound state that crosses the Fermi level as any of the phases are varied. (Indeed, a single scalar equation, $|\bm{S}|=|\bm{T}|$, determines the position of this crossing.)

More interestingly for our purpose, the states $|\sigma+\rangle$ and $|\sigma-\rangle$ cross each other when $\bm{T}=0$. On the other hand, at $\bm{S}=0$, the crossing is between the states $|\sigma+\rangle$ and $|\bar\sigma-\rangle$, where $\bar\sigma=-\sigma$. Each such Majorana-Weyl crossing is determined by three scalar equations. Therefore they generally occur at isolated points in the three-dimensional space of phase differences of the four-terminal junction. In the specific case where all $\phi_{ab}=0$, the Weyl points determined by $\bm{S}=0$ and $\bm{T}=0$ take place at zero energy and coincide at the same set of phases, $(\chi_1,\chi_2,\chi_3)=(0,0,0)\,\mathrm{mod}\,2\pi$. But, in general, the Weyl points do not coincide and occur at finite energy. Due to particle-hole symmetry, the Weyl crossings at a given value of the phases $(\chi_1^*,\chi_2^*,\chi_3^*)$ occuring at energies $\pm E^*$ carry the same topological charge. Furthermore, $2\pi$-phase translations~\cite{footnote} bring another set of two Weyl points with the same charge. Thus, eight pairs of Weyl points at opposite energies with the same topological charges appear in the region $0<\chi_1,\chi_2,\chi_3<4 \pi$ at phases $(\chi_1^*+2\pi n_1,\chi_2^*+2\pi n_2,\chi_3^*+2\pi n_3)$ with $n_i=0,1$. The fermion doubling theorem~\cite{fermion-doubling} ensures that eight other pairs of Weyl points with the opposite topological charges must exist in the same region of phases.

\begin{figure}
(a) 
\includegraphics[width=0.4\columnwidth]{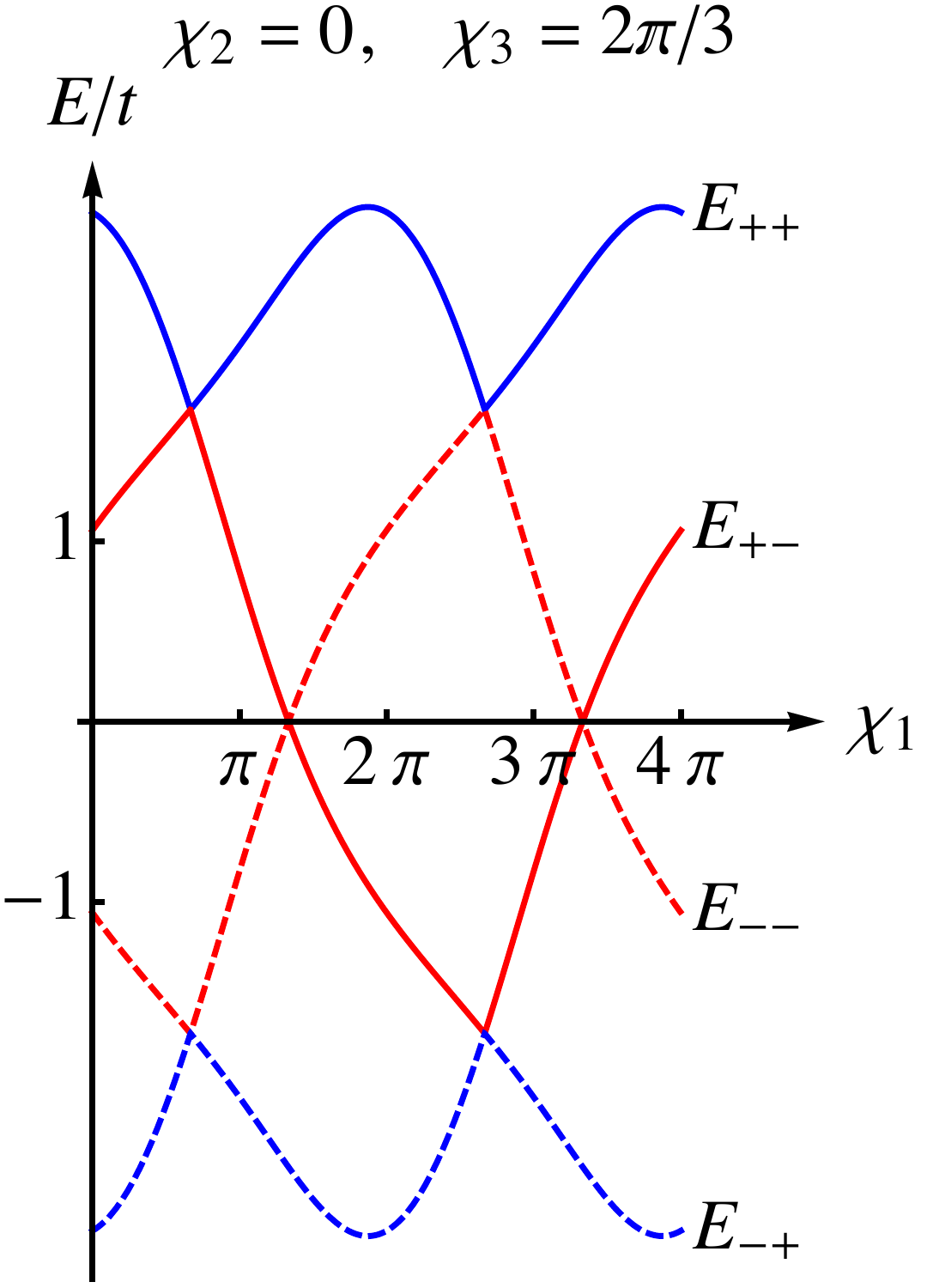}
(b)
\includegraphics[width=0.4\columnwidth]{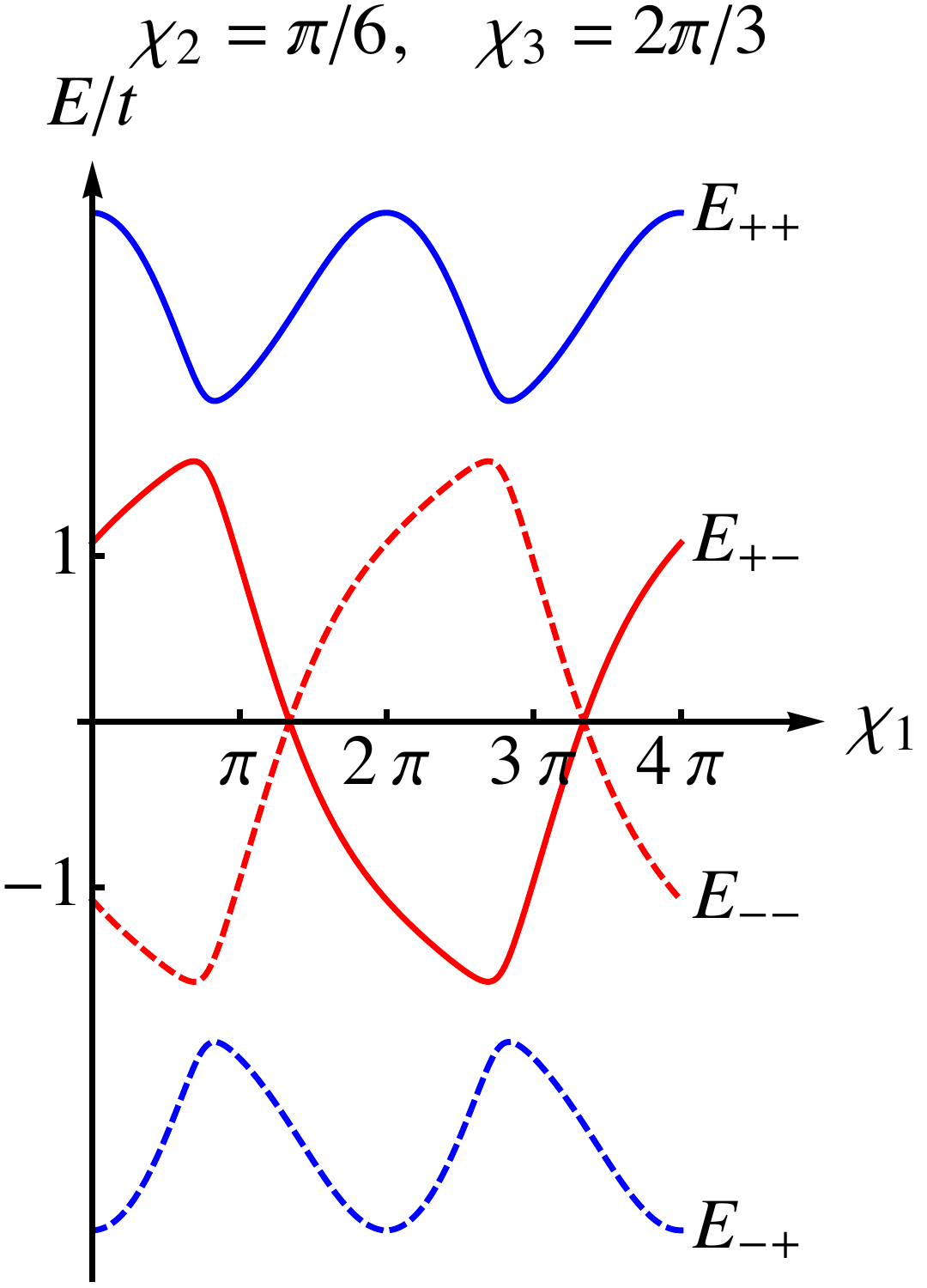}
\\
(c)
\includegraphics[width=0.4\columnwidth]{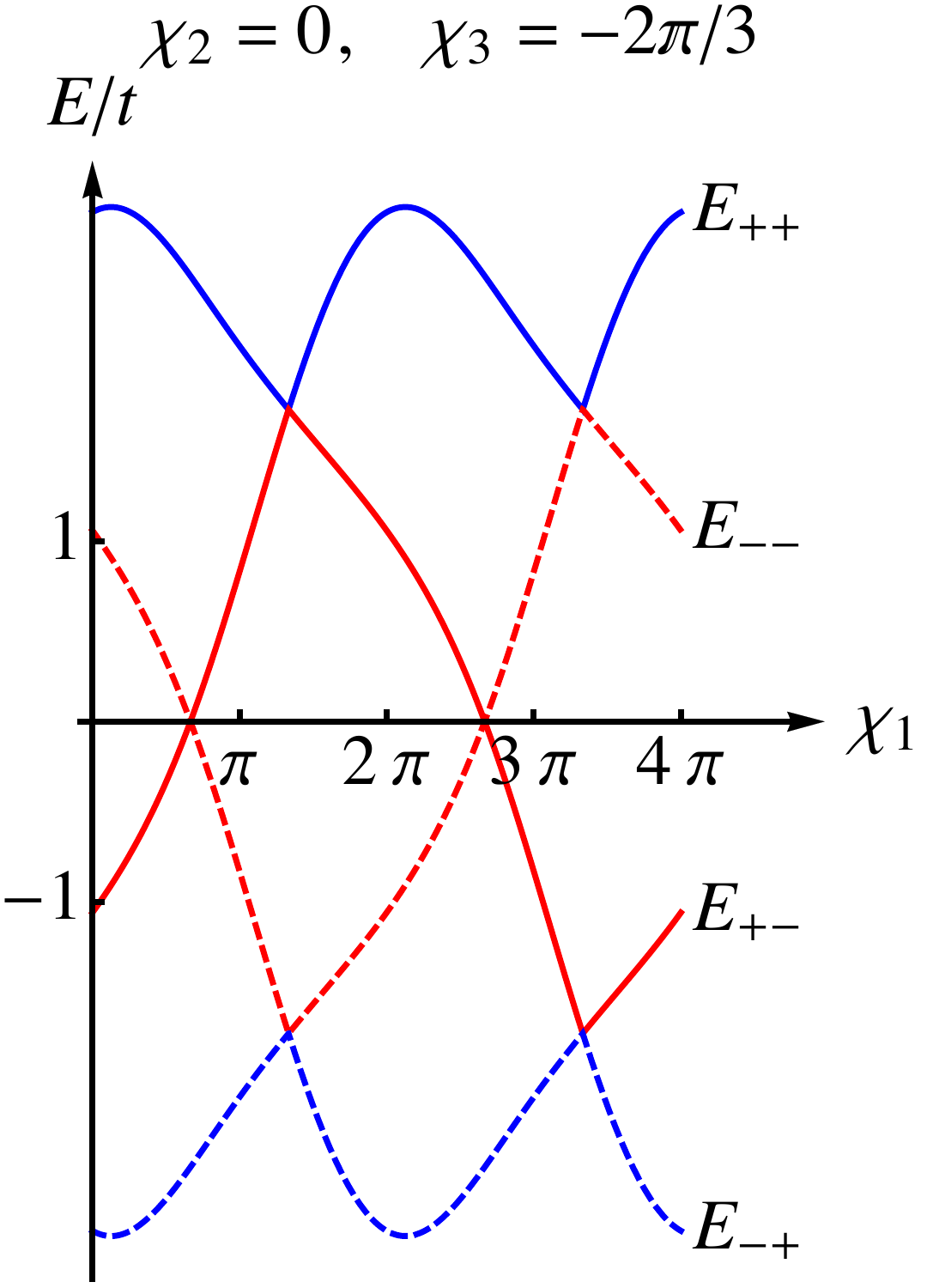}
(d)
\includegraphics[width=0.35\columnwidth]{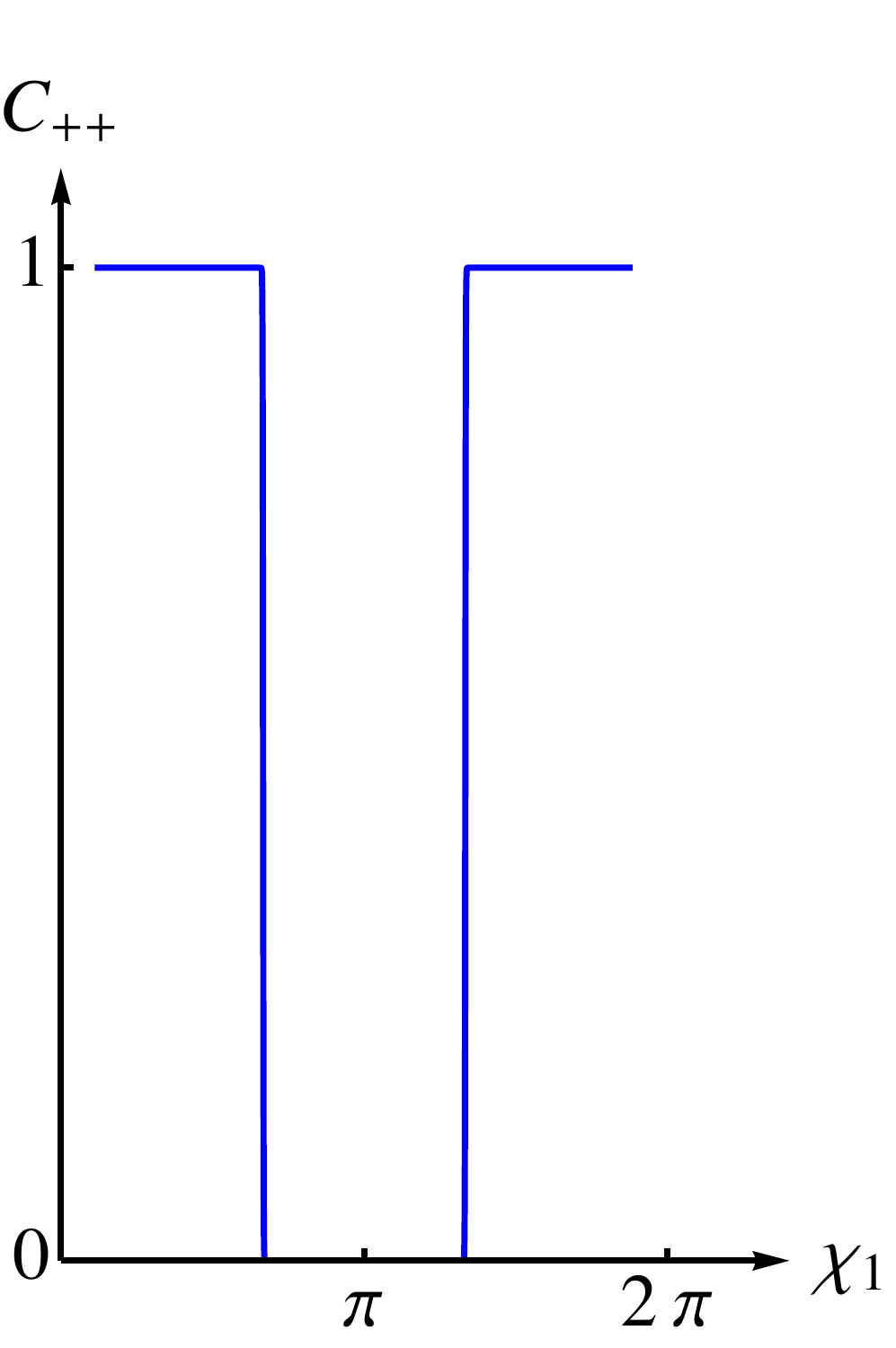}
\caption{
\label{fig:spectra} 
Phase dependence of (a)-(c) the Andreev spectrum and (d) the Chern number in a symmetric four-terminal tunnel junction with $t'/t=0.2$ and $\phi=4\pi/3$.
}
\end{figure}

We show typical spectra in Fig.~\ref{fig:spectra} for a symmetric junction with $t_{aa\pm1}=te^{\pm i\phi/4}$ and $t_{aa\pm2}=t'$ with $t,t',\phi$ real. The Weyl crossings at $\bm{T}=0$ and $\bm{S}=0$ take place at phases $(\phi/2,0,\phi/2)$ and $(-\phi/2,0,-\phi/2)\,\mathrm{mod}\,2\pi$, as illustrated in Figs.~\ref{fig:spectra}(a) and \ref{fig:spectra}(c), respectively; Fig.~\ref{fig:spectra}(b) shows a gapped Andreev spectrum.

Weyl points are monopoles for Berry curvature. Fixing the phase $\chi_1$, we define the Berry curvatures 
\begin{equation}
B_{\sigma\tau}(\chi_1;\chi_2,\chi_3)=-2\,\mathrm{Im}\left\{(\partial_{\chi_2}\langle \sigma\tau|)\partial_{\chi_3}|\sigma\tau\rangle\right\}
\end{equation}
 in the $(\chi_2,\chi_3)$-plane of the two remaining phase differences for each state $|\sigma\tau\rangle$.
 Integration over the region $0<\chi_2,\chi_3<4\pi$ then yields the (quantized) first Chern numbers,
\begin{equation}
C_{\sigma\tau}(\chi_1)=\frac{1}{8\pi}\int_0^{4\pi}\!\!\!d\chi_2\int_0^{4\pi}\!\!\!d\chi_3\;
B_{\sigma\tau}(\chi_1;\chi_2,\chi_3).
\end{equation}
Possible values are constrained by symmetry considerations. Namely, particle-hole symmetry imposes $C_{+\tau}(\chi_1)=-C_{-\tau}(\chi_1)$. Moreover, while the states $|\sigma+\rangle$ are $2\pi$-periodic, shifting one of the phases by $2\pi$ exchanges the states $|+-\rangle$ and $|--\rangle$. Therefore, the latter two  states  have to carry the same Chern number. Together with particle-hole symmetry, this imposes $C_{\sigma-}(\chi_1)=0$.

As the phase $\chi_1$ is varied across the Weyl point at $\chi_1^*$, $C_{\sigma+}(\chi_1)$ changes by $-\sigma Q^*$, where $Q^*$ is the topological charge of the Weyl point, see Fig.~\ref{fig:spectra}(d) for an illustration. The fact, that $C_{\sigma-}(\chi_1)$ remains zero can be understood from the observation that the states $|\sigma-\rangle$ participate in two Weyl crossings (with the state $|-+\rangle$) as the higher energy state and in two other Weyl crossings (with the state $|++\rangle$) as the lower energy state, such that the different contributions to the Berry curvature cancel each other.

We are now in a position to compute the currents through the junction. As the BdG formalism describing superconducting heterostructures introduces a double counting of states, only two of the four states are physical. We choose to keep the states $\sigma=+$.
According to \cite{Riwar2016} (see also Supplemental Material~\cite{SM}), in a multi-terminal Josephson junction, the Andreev states' Berry curvatures determine a non-adiabatic correction to the Josephson currents flowing through two voltage-biased terminals, 
\begin{equation}
\label{eq:Iad}
I_{2,3}(t)=2e\sum_{\tau}\left[ \frac{1}\hbar \frac{\partial E_{+\tau}}{\partial \chi_{2,3}}
\mp \tau \dot\chi_{3,2} B_{+\tau}(\chi_1;\chi_2,\chi_3)\right] h_\tau(t),
\end{equation}
where $\dot\chi_{2,3}=2eV_{2,3}/\hbar$ with dc voltage biases $V_2$ and $V_3$ in terminals 2 and 3, and we used $B_{+\tau}(\chi_1;\chi_3,\chi_2)=-B_{+\tau}(\chi_1;\chi_2,\chi_3)$.
Furthermore $h_\tau=n_\tau-1/2$ describes the (time-dependent) occupations $n_\tau$ of the states $|+\tau\rangle$. 

At fixed occupation of the states, we find that the time-averaged currents are given as
\begin{equation}
%\label{eq:Iad}
\bar I_{2,3}=\mp \frac{2e^2}hV_{3,2} C_{++}(\chi_1) h_+.
\end{equation}
As the state $|++\rangle$ lies above the Fermi level, we may assume $n_+=0$ to obtain the quantized transconductance
\begin{equation}
G_{23}\equiv\frac{\bar I_2}{V_3}=\frac{2e^2}h C_{++}(\chi_1)
,\qquad G_{32}\equiv\frac{\bar I_3}{V_2}=-G_{23}.\label{eq-main}
\end{equation}
The unit of transconductance quantization is half the one found in \cite{Riwar2016} because of the lifted spin degeneracy in the junction considered here.

Equation \eqref{eq-main} is our main result. While it ressembles the predicted transconductance quantization in non-topological four-terminal junctions (taken apart the modified unit of transconductance quantization), there are important differences. As the Andreev spectrum does not have a gap at the Fermi level, the transconductance does not probe the ground state of the system. While the $4\pi$-periodic state $|+-\rangle$ does not carry a Chern number itself, it is essential in transferring Berry curvature across the Fermi level. Similarly to signatures of Majorana physics in two-terminal junctions~\cite{Fu2009}, the observation of conductance quantization requires that the system does not relax to its equilibrium occupations~\cite{footnote-local}. As the result does not depend on the occupation $n_-$ of the state $|+-\rangle$, however, it is robust with respect to random switchings.

{\em Arbitrary junction.} Our result is not restricted to tunnel junctions.  In general, we may use the formalism of \cite{Beenakker1991} to find that the Andreev spectrum is determined by
\begin{equation}
\label{eq:det}
\mathrm{det}\left[1+a^2(E) S(E) e^{i\chi}S^*(-E)e^{-i\chi}\right]=0.
\end{equation}
Here $S(E)$ is the $4\times 4$ scattering (or $S$-)matrix for electrons with energy $E$ between the four one-dimensional leads, $S^*(-E)$ is the corresponding $S$-matrix for holes, $\chi$ is a diagonal matrix whose diagonal elements $(\chi_1,\chi_2,\chi_3,\chi_4=0)$ are the superconducting phases, and $a(E)=E/\Delta-i\sqrt{1-(E/\Delta)^2}$ is the Andreev reflection amplitude. The important difference between Eq.~\eqref{eq:det} and a similar one used in \cite{Riwar2016} is the reversed sign in front of the second term in the determinant. It originates from the $\pi$-phase shift in the Andreev reflection processes between electrons and holes incident upon a $p$-wave superconductor, in contrast with the $s$-wave case considered earlier.

As it was noticed in \cite{Yokoyama2015}, $a^2(E)=-a^2(\sqrt{\Delta^2-E^2})$. Therefore, the solutions of Eq.~\eqref{eq:det} near the gap edge can be related with those found in \cite{Riwar2016} near the Fermi level, and vice-versa. In particular, in the short-junction limit in which the energy dependence of the normal $S$-matrix (on the scale of Thouless energy $E_T\gg\Delta$) can be neglected, $S(E)\approx S(0)$, we readily find that {\em (i)} the state $|++\rangle$ has finite probability~\cite{footnoteRMT,footnote2} to merge with the continuum spectrum at isolated points in the phase space (the equivalents of Andreev-Weyl crossings at zero energy found in \cite{Riwar2016}), {\em (ii)} state $|+-\rangle$ crosses the Fermi level along surfaces in the phase space (the equivalent of states merging with the continuum at the gap edge in the Supplementary Information of \cite{Riwar2016}), and {\em (iii)} the four Andreev states cross each other at zero energy and phases $\chi_1=\chi_2=\chi_3=0$ in the time-reversal symmetric case, $S(0)= S^T(0)$. In the latter case, two Weyl crossings with opposite charges are superposed and $C_{++}(\chi_1)=0$ at any $\chi_1$.

The possibility of Weyl crossings at finite energy was also mentioned in the Supplementary Information of\cite{Riwar2016}, in which context they were playing no role in the transconductance quantization. 
We can characterize their occurence using random matrix theory~\cite{footnoteRMT}, namely by drawing scattering matrices from the circular unitary ensemble to describe systems without time-reversal symmetry. We find that the total probability to realize Majorana-Weyl crossings is $82$\%. The probability density for them to occur at a given energy is shown in Fig.~3.  (There are $14$\% matrices with both Majorana-Weyl and gap-edge touchings in their Andreev spectrum.) 

\begin{figure}
(a)
\includegraphics[width=0.42\columnwidth]{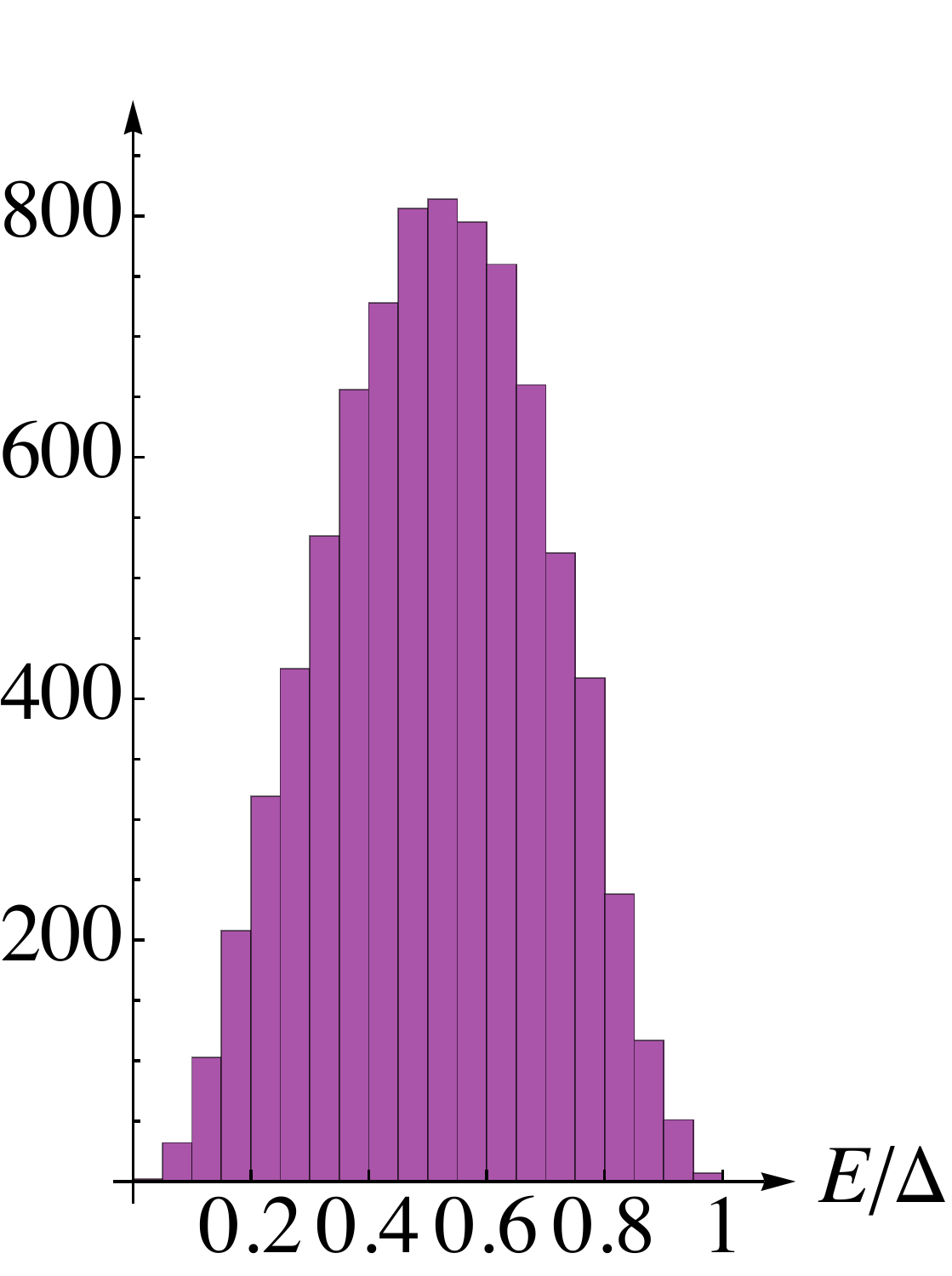}
\,(b)
\includegraphics[width=0.42\columnwidth]{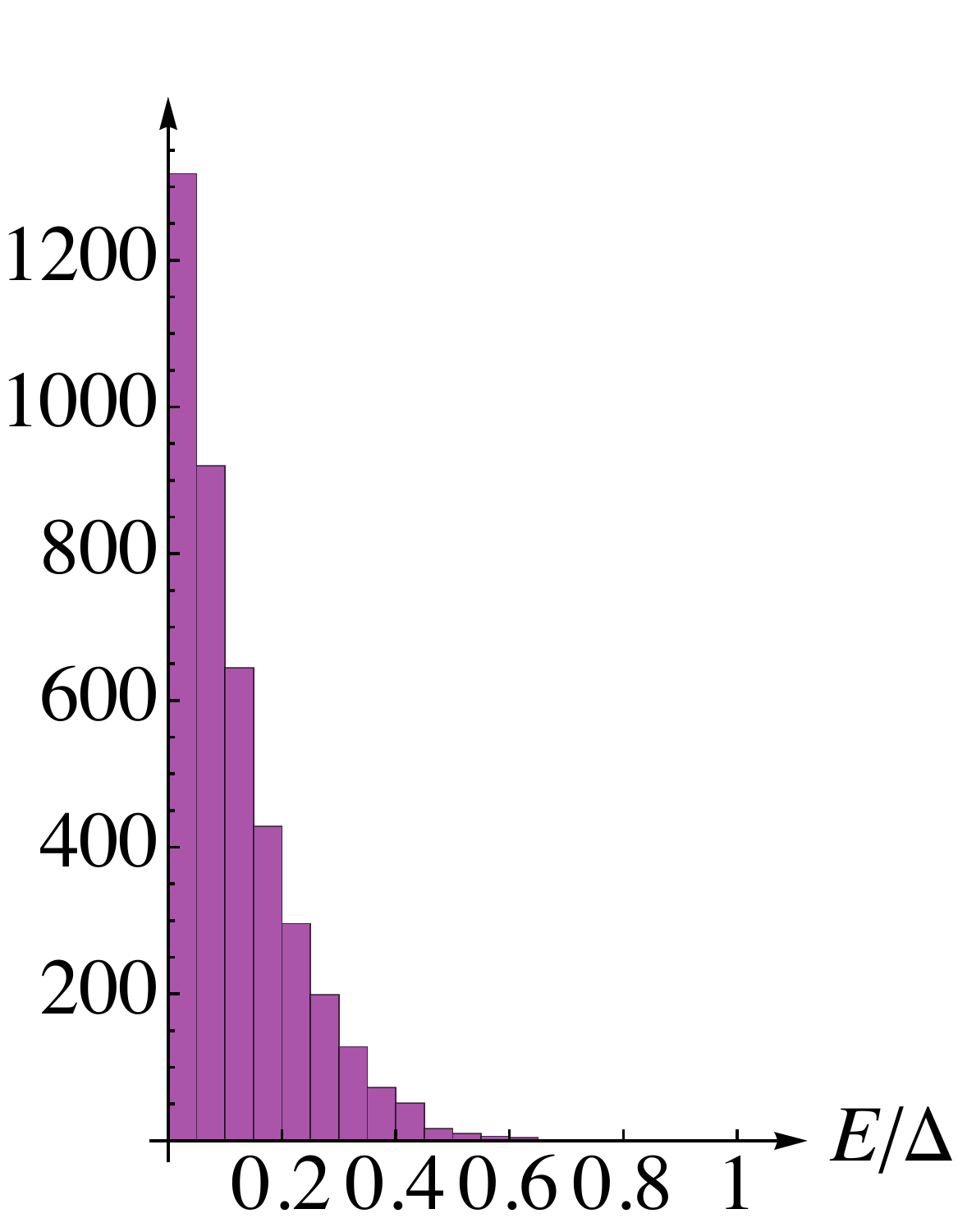}
\caption{
\label{fig:hisogram} 
Histogram of (a) the energies and (b) energy difference (both in absolute value) of two Majorana-Weyl crossings in $5\,000$ short, four-terminal Josephson junctions through a normal region described by a scattering matrix drawn out of the circular unitary ensemble (no time-reversal symmetry).}
\end{figure} 

Note that the tunnel case discussed earlier can be recast within the scattering formalism, where it corresponds to a normal state scattering matrix
\begin{equation}
S=(1-i\pi\nu  T)^{-1}(1+i\pi\nu  T),
\end{equation}
where $T$ is the matrix of tunnel hopping elements between the leads and $\nu$ is the normal density of states. Indeed, using $T=T^\dagger$ with $|T_{ab}|\ll \nu^{-1}$ and $a(E)\approx E/\Delta-i$ at $|E|\ll\Delta$, we may recast Eq.~\eqref{eq:det} as a Hamiltonian equation
\begin{equation}
\label{eq:BdG-bis}
E\psi_a=2i\pi\nu\Delta \sum_b |T_{ab}|\sin\left(\frac{\chi_a-\chi_b}2-\phi_{ab}\right)\psi_b
\end{equation}
with $T_{ab}= |T_{ab}|e^{i\phi_{ab}}$. The corresponding Hamiltonian is equivalent to Eq.~\eqref{eq:Htunnel} provided one identifies $|t_{ab}|=\frac12\sqrt{{\cal T}_{ab}}\Delta$, where ${\cal T}_{ab}=4\pi^2\nu^2|T_{ab}|^2\ll 1$ is the transmission probability between leads $a$ and $b$.

{\em Experimental realizations.} The model studied above is applicable to junctions made with crossed nanowires like in \cite{Gazibegovic2017}. Alternatively, a four-terminal Josephson junction can be realized by depositing superconductors on the edges of a quantum point contact made with a quantum spin-Hall insulator. In the presence of time-reversal symmetry, back-scattering within a single edge is forbidden. In that case, we find that the $2\pi$- and $4\pi$-periodic Andreev states become degenerate along lines in the space of phases rather than at isolated points. If time-reversal symmetry is broken, we recover the results for the crossed nanowires discussed before.

{\em Conclusions.} In this work we unveiled a topological property of the Andreev spectrum in multiterminal junctions between topological superconductors. Namely, we predicted that finite-energy Weyl crossings between $2\pi$- and $4\pi$-periodic Andreev states may result in a quantized transconductance in units of $2e^2/h$ between two voltage-bias leads. We anticipate the conditions for the robustness of this prediction in the presence of non-adiabatic effects will be different from the case of conventional superconductors~\cite{Eriksson2017}. Furthermore, it would be interesting to understand whether the result found in this work can be analyzed within the general classification of topological insulators and superconductors~\cite{Chiu2016,Shiozaki2014}.

\acknowledgments
We thank Y.~Oreg for urging us to address this problem, as well as Y.~Nazarov for interesting discussion. We acknowledge funding by the ANR through the grant  ANR-17-PIRE-0001.

\onecolumngrid
\pagebreak
\clearpage

\setcounter{equation}{0}
\setcounter{figure}{0}
\setcounter{table}{0}
\setcounter{page}{1}
\renewcommand{\thefigure}{S\arabic{figure}}
\renewcommand{\theequation}{S\arabic{equation}}
\renewcommand{\citenumfont}[1]{S#1}
\renewcommand{\bibnumfmt}[1]{[S#1]}
\begin{center}
\textbf{\large Supplemental Material: \\
``Majorana-Weyl crossings in topological multi-terminal junctions" }
\end{center}

\twocolumngrid

Below we derive Eq. (7) in the main text for the non-adiabatic correction to the Josephson currents flowing in a multi-terminal junction between spinless $p$-wave superconducting leads in the presence of a voltage bias.

{\bf 1. } Without voltage bias, the (second-quantized) Hamiltonian describing a multi-terminal junction with spinless $p$-wave superconducting leads attached through a normal scattering region can be put in the form
\begin{equation}
H=\frac 12 \Psi^\dagger{\cal H}\Psi.
\end{equation}
Here $\cal H$ is a BdG (first-quantized) Hamiltonian and the electron annihilation and creation operators are gathered in a Nambu annihilation operator
\begin{equation}
\label{eq-SM:Nambu}
\Psi=\left(\begin{array}{c}c\\c^\dagger
\end{array}
\right).
\end{equation}
The BdG Hamiltonian defines an eigenproblem,
\begin{equation}
\label{eq-SM:Phi}
{\cal H}\left(\begin{array}{c}u_n\\v_n
\end{array}
\right)=\varepsilon_n\left(\begin{array}{c}u_n\\v_n
\end{array}
\right),
\end{equation}
with the normalization condition
\begin{equation}
\int dx \left[u_n(x)u^*_m(x)+v_n(x)v^*_m(x)\right]=\delta_{nm}
\end{equation}
or, equivalently,
\begin{eqnarray}
{\sum_n} u_n(x)u^*_n(y)={\sum_n}v_n(x)v^*_n(y)=\delta(x-y)
\nonumber \\ 
\mathrm{and}\qquad {\sum_n} u_n(x)v^*_n(y)=0.
\end{eqnarray}
The solutions possess particle-hole symmetry. Indeed, if $\varepsilon_n$ is an eigenenergy associated with an eigenvector $\Phi_n=(u_n,v_n)^T$, then $-\varepsilon_n$ is another eigenenergy associated with the eigenvector $\tilde\Phi_n=(v_n^*,u_n^*)^T$.
Using these solutions and 
\begin{equation}
\label{eq-SM:Psi}
\Psi={\sum_n}'\left[\left(\begin{array}{c}u_n\\v_n
\end{array}\right)\gamma_{n}+\left(\begin{array}{c}v_n^*\\u_n^*
\end{array}\right)\gamma^\dagger_{n}\right],
\end{equation}
where the prime indicates that the sum is restricted to {\it only one} of the two particle-hole symmetric states, we can diagonalize the Hamiltonian as
\begin{equation}
H={\sum_{n}}'\varepsilon_n\gamma^\dagger_{n}\gamma_{n}.
\end{equation}
The restricted sum is necessary to ensure Fermi commutation relations for the operators $\gamma_{n}$ \cite{book-SM}.
%, and is sometimes referred as the way to resolve double-counting introduced by the Nambu operator \eqref{eq-SM:Nambu} \cite{double-counting}. {\cc{\tt not sure what you want to say .. it does resolve the problem, no ?}} 
Note that it does not matter whether the state with $\varepsilon_n>0$ or $\varepsilon_n<0$ is retained in the sum. However, for later convenience, we will retain a state whose wave function depends continuously on the superconducting phases.

{\bf 2. }We may similarly express the current operator in lead $k$
\begin{equation}
\label{eq-SM:I}
I_k=\frac 12\Psi^\dagger{\cal I}_k\Psi,
\end{equation}
where ${\cal I}_k=(2e/\hbar)\partial {\cal H}/\partial{\chi_k}$ and $\chi_k$ is the (fixed) superconducting phase in lead $k$. Inserting Eq.~\eqref{eq-SM:Psi} into \eqref{eq-SM:I} and defining occupations $f_{n}=\langle \gamma^\dagger_{n}\gamma_{n}\rangle$, we get for the current expectation value
\begin{equation}
\langle I_k\rangle=\frac{e}\hbar{\sum_n}'\left[f_{n}\int\Phi_n^\dagger\frac{\partial {\cal H}}{\partial{\chi_k}}\Phi_n+(1-f_{n})\int\tilde\Phi_n^\dagger\frac{\partial {\cal H}}{\partial{\chi_k}}\tilde\Phi_n\right]
%\nonumber\\
%&=&{\cc{ -\frac{2e}\hbar{\sum_n}'\left(\frac12-f_{n}\right)\int\Phi_n^\dagger\frac{\partial {\cal H}}{\partial{\chi_k}}\Phi_n}}
.\label{eq-SM:Iexp}
\end{equation}
(Note that we use notation $\int \phi\equiv\int dx\; \phi(x)$.)

{\bf 3. }Following \cite{Thouless1983-SM}, we now look for an adiabatic expansion of the solution of the BdG Hamiltonian in the presence of slowly time-varying phases $\chi_k(t)$,
\begin{equation}
\label{eq-SM:BdG}
{\cal H}[{\bm \chi}(t)]\Phi
=i\hbar\frac{\partial}{\partial t}\Phi,
\end{equation}
in the form $\Phi(t)=\sum_nc_n(t)\Phi_n[{\bm \chi}(t)]$ with ${\bm \chi}(t)=\{{ \chi}_k(t)\}$. Note that there is no prime in the sum as we expand $\Phi$ in the complete basis of adiabatic solutions of Eq.~\eqref{eq-SM:Phi} at each set of phases ${\bm \chi}$. Then, Eq.~\eqref{eq-SM:BdG} reads equivalently
\begin{equation}
\label{eq-SM:BdG2}
i\hbar\dot c_n-\varepsilon_n c_n=-i\hbar\sum_{mk}\dot\chi_kc_m\left(\int \Phi_n^\dagger\frac{\partial \Phi_m}{\partial \chi_k}\right).
\end{equation}
Taking the initial condition $c_p(t=0)=\delta_{pn}$ to determine the adiabatic expansion for the state $\Phi^{(n)}$, we find that Eq.~\eqref{eq-SM:BdG2} yields in leading order
\begin{equation}
\label{eq-SM:BdG3}
i\hbar\dot c_n-\left[
\varepsilon_n -i\hbar\sum_{k}\dot\chi_k
\left(\int \Phi_n^\dagger\frac{\partial \Phi_n}{\partial \chi_k}\right)
\right]
c_n=0.
\end{equation}
The solution is given as $c(t)=\exp[i\theta(t)]$ with
\begin{equation}
\label{eq-SM:cn}
\theta(t)=
-\frac 1\hbar\int\limits_0^tds\;\varepsilon_n({\bm \chi(s)})+\sum_k\int\limits_{\chi(0)}^{\chi(t)} d\chi_k\;{\cal A}_{n,k}({\bm \chi(s)}),
\end{equation}
where ${\cal A}_{n,k}=i\int \Phi_n^\dagger({\partial \Phi_n}/{\partial \chi_k})$ is the (real) Berry connection.

In the next order, the (small) coefficients $c_{m\neq n}(t)$ satisfy the equation
\begin{equation}
\label{eq-SM:BdG4}
i\hbar\dot c_m-
\varepsilon_m c_m
=
 -i\hbar\sum_{k}\dot\chi_k
\left(\int \Phi_m^\dagger\frac{\partial \Phi_n}{\partial \chi_k}\right)
e^{i\theta(t)}.
\end{equation}
Neglecting the small Berry connection contributions,  the solution reads
\begin{equation}
\label{eq-SM:BdG4}
c_m\approx
 -\frac {i\hbar}{\varepsilon_n-\varepsilon_m}\sum_{k}\dot\chi_k
\left(\int \Phi_m^\dagger\frac{\partial \Phi_n}{\partial \chi_k}\right)
e^{i\theta(t)}.
\end{equation}
Combining these results, we obtain the eigenvectors $\Phi^{(n)}(t)$  in leading order in $\dot \chi_k$:
\begin{widetext}
\begin{equation}
\label{eq-SM:WF}
\Phi^{(n)}(t)=e^{i\theta(t)}\left[
\Phi_n({\bm \chi}(t))-i\hbar\sum_k\dot \chi_k\sum_{m\neq n}
\frac 1{\varepsilon_n({\bm \chi}(t))-\varepsilon_m({\bm \chi}(t))}
\left(\int \Phi_m^\dagger\frac{\partial \Phi_n}{\partial \chi_k}\right)\Phi_m({\bm \chi}(t))
\right].
\end{equation}

{\bf 4. } The instantaneous current is obtained by replacing $\Phi_n$ with $\Phi^{(n)}$ in Eq.~\eqref{eq-SM:Iexp}. Using Eq.~\eqref{eq-SM:WF} and standard properties of eigenstates, we  evaluate 
\begin{eqnarray}
\int\Phi^{(n)\dagger}\frac{\partial {\cal H}}{\partial{\chi_k}}\Phi^{(n)}
&=&
\int\Phi_n^\dagger\frac{\partial {\cal H}}{\partial{\chi_k}}\Phi_n
\\&&-i\hbar\sum_l\dot \chi_l\sum_{m\neq n}
\frac 1{\varepsilon_n-\varepsilon_m}
\left[
\left(\int\Phi_n^\dagger\frac{\partial {\cal H}}{\partial{\chi_k}}\Phi_m\right)
\left(\int \Phi_m^\dagger\frac{\partial \Phi_n}{\partial \chi_l}\right)
\right.
\left.-
\left(\int\Phi_m^\dagger\frac{\partial {\cal H}}{\partial{\chi_k}}\Phi_n\right)
\left(\int \frac{\partial \Phi^\dagger_n}{\partial \chi_l}\Phi_m\right)
\right]
\nonumber
\\
&=&
\frac{\partial \varepsilon_n}{\partial{\chi_k}}
+2\hbar\,\mathrm{Im}\sum_l\dot \chi_l\sum_{m\neq n}
\frac 1{\varepsilon_n-\varepsilon_m}
\left(\int\Phi_n^\dagger\frac{\partial {\cal H}}{\partial{\chi_k}}\Phi_m\right)
\left(\int \Phi_m^\dagger\frac{\partial \Phi_n}{\partial \chi_l}\right)
\nonumber
\\
&=&
\frac{\partial \varepsilon_n}{\partial{\chi_k}}
+2\hbar\sum_l\dot \chi_l\,\mathrm{Im}\sum_{m\neq n}
\left(\int\frac{\partial \Phi_n^\dagger}{\partial{\chi_k}}\Phi_m\right)
\left(\int \Phi_m^\dagger\frac{\partial \Phi_n}{\partial \chi_l}\right)
\nonumber
\\
&=&
\frac{\partial \varepsilon_n}{\partial{\chi_k}}
+2\hbar\sum_l\dot \chi_l\,\mathrm{Im}
\left(\int\frac{\partial \Phi_n^\dagger}{\partial{\chi_k}}\frac{\partial \Phi_n}{\partial \chi_l}\right)
=\frac{\partial \varepsilon_n({\bm \chi}(t))}{\partial{\chi_k}}
-\hbar\sum_l\dot \chi_l{\cal B}_{n,kl}({\bm \chi}(t)),
\end{eqnarray}
\end{widetext}
where ${\cal B}_{n,kl}=\partial_k{\cal A}_{n,l}-\partial_l{\cal A}_{n,k}$ is the Berry curvature of level $n$, up to first order in $\dot{\bm\chi}$.

We readily check that the corresponding expression for the particle-hole conjugated states has the opposite sign, $\int{\tilde\Phi}^{(n)\dagger}\left(\partial{\cal H}/\partial\chi_k\right)\tilde\Phi^{(n)}=-\int\Phi^{(n)\dagger}\left(\partial{\cal H}/\partial\chi_k\right)\Phi^{(n)}$. Then, the average current in lead $k$ is given as
\begin{equation}
\label{eq-supl:Itot}
\langle I_k\rangle={\sum_n}' (\frac 12-f_{n})
\left[
-\frac{2e}\hbar\frac{\partial \varepsilon_n}{\partial{\chi_k}}
+2e\sum_l\dot \chi_l{\cal B}_{n,kl}
\right].
\end{equation}
This result is identical to the one derived in \cite{Riwar2016-SM} for multi-terminal junctions with conventional superconducting leads. In the latter case, a summation over spins yields the additional factor 2 in the unit of conductance quantization.

%\end{appendix}
%\end{widetext}


\begin{thebibliography}{99}
\bibitem{footnote}
Here we used $S_x(\chi_1+2\pi,\chi_2,\chi_3)=T_x(\chi_1,\chi_2,\chi_3)$ and $S_{y,z}(\chi_1+2\pi,\chi_2,\chi_3)=-T_{y,z}(\chi_1,\chi_2,\chi_3)$, and similar relations for other translations.
\bibitem{footnote-local} In equilibrium, the local Berry curvature of the $4\pi$-periodic state will compensate the contribution of the $2\pi$-periodic state.
\bibitem{footnoteRMT}
The matrix $A=Se^{i\chi}S^*e^{-i\chi}$ admits four pairwise-conjugated eigenvalues on the unit circle. Using this property, we find that the condition for $A$ to admit $-1$ as an eigenvalue, and an Andreev state to touch the continuum (Andreev-Weyl crossing), is the existence of a set of phases such that
\[
4+4\, \text{Tr} A+( \text{Tr} A)^2- \text{Tr} (A^2)=0.
\]
Furthermore, the condition for an eigenvalue to be doubly degenerate, and thus a Majorana-Weyl crossing to occur, is 
\[
8-( \text{Tr} A)^2+4\, \text{Tr} (A^2)=0.
\]
The energy at the crossing is $\pm\sqrt{(4-\text{Tr} A)/8}$.
\end{thebibliography}
\end{document}